\documentclass[12pt,preprint]{aastex}

\newcommand{\etal }{{et al.} }
\newcommand{\msun}{\thinspace M_\odot}

\def\lesssim{\mathrel{\hbox{\rlap{\hbox{\lower4pt\hbox{$\sim$}}}\hbox{$<$}}}}
\def\gtrsim{\mathrel{\hbox{\rlap{\hbox{\lower4pt\hbox{$\sim$}}}\hbox{$>$}}}}
\newcommand{\cm}{\,{\rm cm}^{-3} } 
\newcommand{\km}{\,{\rm km\, s}^{-1}} 
\newcommand{\nc}{n_{\rm c} }

\newcommand{\dfrac}[2]{{\displaystyle \frac{#1}{#2}} }

\shorttitle{Population III Star Formation}
\shortauthors{Machida 2010}

\begin{document}
\title{Population III Star Formation in Magnetized Primordial Clouds}
\author{Masahiro N. Machida\altaffilmark{1}} 
\altaffiltext{1}{National Astronomical Observatory of Japan, Mitaka, Tokyo 181-8588; masahiro.machida@nao.ac.jp}

\begin{abstract}
The evolution of primordial collapsing clouds and formation of proto-Population III stars are investigated using three-dimensional ideal MHD simulations.
We calculated the collapse of magnetized primordial clouds from the prestellar stage until the epoch after the proto-Population III star formation, spatially resolving both parsec-scale clouds and sub-AU scale protostars.
The formation process of proto-population III star is characterized by the ratio of rotational to magnetic energy of the natal cloud.
When the rotational energy is larger than the magnetic energy, fragmentation occurs in the collapsing primordial cloud before the proto-Population III star formation and binary or multiple system appears.
Instead, when the magnetic energy is larger than the rotational energy, strong jet with $>100\,{\rm km\,s}^{-1}$ is driven by circumstellar disk around the proto-population III star without fragmentation.
Thus, even in the early universe, the magnetic field plays an important role in the star formation process.
\end{abstract}

\section{Introduction}
Magnetic fields is a key ingredient in present-day star formation.
For example, protostellar jets, which are ubiquitous in star-forming regions, are considered to be driven from protostars by the Lorentz force.
Protostellar jets influence gas accretion onto protostars and disturb the ambient medium.
In addition, the angular momentum of the cloud is removed by magnetic braking and protostellar jets.
The removal of angular momentum makes protostar formation possible in a parent cloud that has a much larger specific angular momentum than the protostar.
So far, magnetic effects in primordial gas clouds have been ignored in many studies because magnetic fields in the early universe are supposed to be extremely weak.
However, recent studies indicate magnetic fields of moderate strength can exist even in the early universe.
Cosmological fluctuations produced magnetic fields before the epoch of recombination \citep{ichiki06}.
These fields were sufficiently large to seed the magnetic fields in galaxies.
A generation mechanism for magnetic fields at the epoch of reionization are also proposed \citep{langer03}, in which  magnetic fields in intergalactic matter are amplified up to $\sim 10^{-11}\ {\rm G}$.
These fields can therefore increase up to $\sim 10^{-7}-10^{-8}\ {\rm G}$ in the first collapsed object having number density of $n \sim 10^3 \cm$.
These fields may influence the evolution of  primordial gas clouds and formation of Population III stars.

Under spherical symmetry including  hydrodynamical radiative transfer, many authors have carefully investigated both the present-day \citep[e.g.,][]{masunaga00} and primordial \citep[e.g.,][]{omukai98} star formation processes.
A significant difference between present-day and primordial star formation exists in the thermal evolution of the collapsing gas cloud because of differences in the abundance of dust grains and metals.
In present-day star formation, the gas temperature in molecular clouds is $\sim10$\,K.
These clouds collapse isothermally for $\nc \lesssim 10^{11}\cm$.
Then the gas becomes adiabatic at $\nc \simeq 10^{11}\cm$, and an adiabatic core (or the first core) forms. 
After the dissociation of molecular hydrogen ($\nc \simeq 10^{16}\cm$), the protostar forms at $\nc \simeq 10^{21}\cm$.
On the other hand, primordial gas clouds have temperatures of $\sim200-300$\,K at $\nc \simeq 10^3\cm$ \citep{omukai05,bromm02}.
These clouds collapse keeping polytropic index $\gamma\simeq 1.1$ for a long range of $10^4\cm \lesssim \nc \lesssim 10^{16}\cm$.
Thus, the first core does not appear in the primordial collapsing cloud.
After the central density reaches $\nc \simeq 10^{16}\cm$, the thermal evolution of the primordial collapsing cloud begins to coincide with that of a present-day cloud  \citep{omukai05}.
The difference in thermal evolution between present-day and primordial clouds arises when $\nc \lesssim 10^{16}\cm$.

Another major difference between present-day and primordial star formation exists in their magnetic evolution.
In present-day star formation, neutral gas is well-coupled  with ions for $\nc \lesssim 10^{12}\cm$ and $\nc \gtrsim 10^{15}\cm$, while the magnetic field dissipates by Ohmic dissipation in the range of $10^{12}\cm \lesssim \nc \lesssim 10^{15}\cm$ \citep{nakano02}. 
In the collapsing cloud, $\sim99$\% of the magnetic field  is dissipated for $10^{12}\cm \lesssim \nc \lesssim 10^{15}\cm$ \citep{machida07}.
On the other hand, in a primordial gas cloud,  the magnetic field couples strongly with the primordial gas during all phases of the star formation, as long as the initial field strength is weaker than $B_0\lesssim 10^{-5}(n/10^3\cm)^{0.55}$\,G \citep{maki04,maki07}.
In summary, the magnetic field is largely dissipated by Ohmic dissipation before protostar formation in present-day clouds, while the magnetic field can continue to be amplified without dissipation in primordial clouds.

The magnetic field in the collapsing cloud is closely related to the fragmentation or formation of binary or multiple stellar systems.  
In present-day star formation, the magnetic field strongly suppresses rotation-driven fragmentation \citep{machida08b}.  
For primordial clouds, magnetic effects on fragmentation are still unknown. 
In this study, we investigate the evolution of weakly magnetized primordial clouds and the formation of Population III stars using three-dimensional simulations.

\section{Model and Numerical Method}
To calculate the evolution of primordial star-forming clouds in a large dynamic range, we use the three-dimensional ideal MHD nested grid method \citep{machida05a,machida05b,machida06a,machida06b}.
For gas pressure, we use a barotropic relation that approximates the result of one-zone calculation where thermal and chemical processes for primordial gas are solved in detail \citep{omukai05}.

As the initial state, we take a spherical cloud whose density is twice higher than that for hydrostatic equilibrium with external pressure (Bonnor-Ebert sphere).
The central (number) density of the Bonnor-Ebert sphere is set at $\rho_{\rm BE}(0) = 3.5\times 10^{-21}\,{\rm g}\cm$.
The initial temperature is set at 250\,K, and the radius of a Bonnor-Ebert sphere is $R_0 =6.5$\,pc.
To promote fragmentation, a small $m=2$-mode density perturbation is imposed on the spherical cloud.
The total mass contained inside $R_0$ is $ M_{\rm c} = 1.8\times 10^4 \msun$.
The initial cloud rotates around the $z$-axis at a uniform angular velocity of $\Omega_0$, and has a uniform magnetic field $B_0$ parallel to the $z$-axis (or rotation axis).
The initial model is characterized by two non-dimensional parameters: the ratios of the rotational energy to the gravitational energy $\beta_0$ ($=E_{\rm rot}/|E_{\rm grav}|$), and of the magnetic energy to the gravitational energy $\gamma_0$ ($=E_{\rm mag}/|E_{\rm grav}|$), where $E_{\rm rot}$, $E_{\rm mag}$ and $E_{\rm grav}$ are the rotational, magnetic and gravitational energies, respectively.
We made 36 models using different values for these two parameters.

We adopted the nested grid method \citep[for details, see ][]{machida05a,machida06a} to obtain high spatial resolution near the center.
Each level of a rectangular grid has the same number of cells ($ = 128 \times 128 \times 64 $), with the cell width $h(l)$ depending on the grid level $l$.
The cell width is halved with every increment of the grid level.
The highest level of the grid changes dynamically: a new finer grid is generated whenever the minimum local Jeans length $\lambda _{\rm J}$ falls below $8\, h (l_{\rm max})$, where $h$ is the cell width. 
The maximum level of grids is restricted to $l_{\rm max} = 30$.
Since the density is highest in the finest grid,  generation of a new grid ensures the Jeans condition with a safety factor of 2.
We begin our calculations with three grid levels ($l=1-3$).
The box size of the initial finest grid $l=3$ is chosen to be $2 R_0$, where $R_0$ is the radius of the critical Bonnor-Ebert sphere. 
The coarsest grid ($l=1$) then has a box size of $2^3\, R_0$.

\section{Results}
\begin{figure}
\begin{center}
\includegraphics[width=140mm]{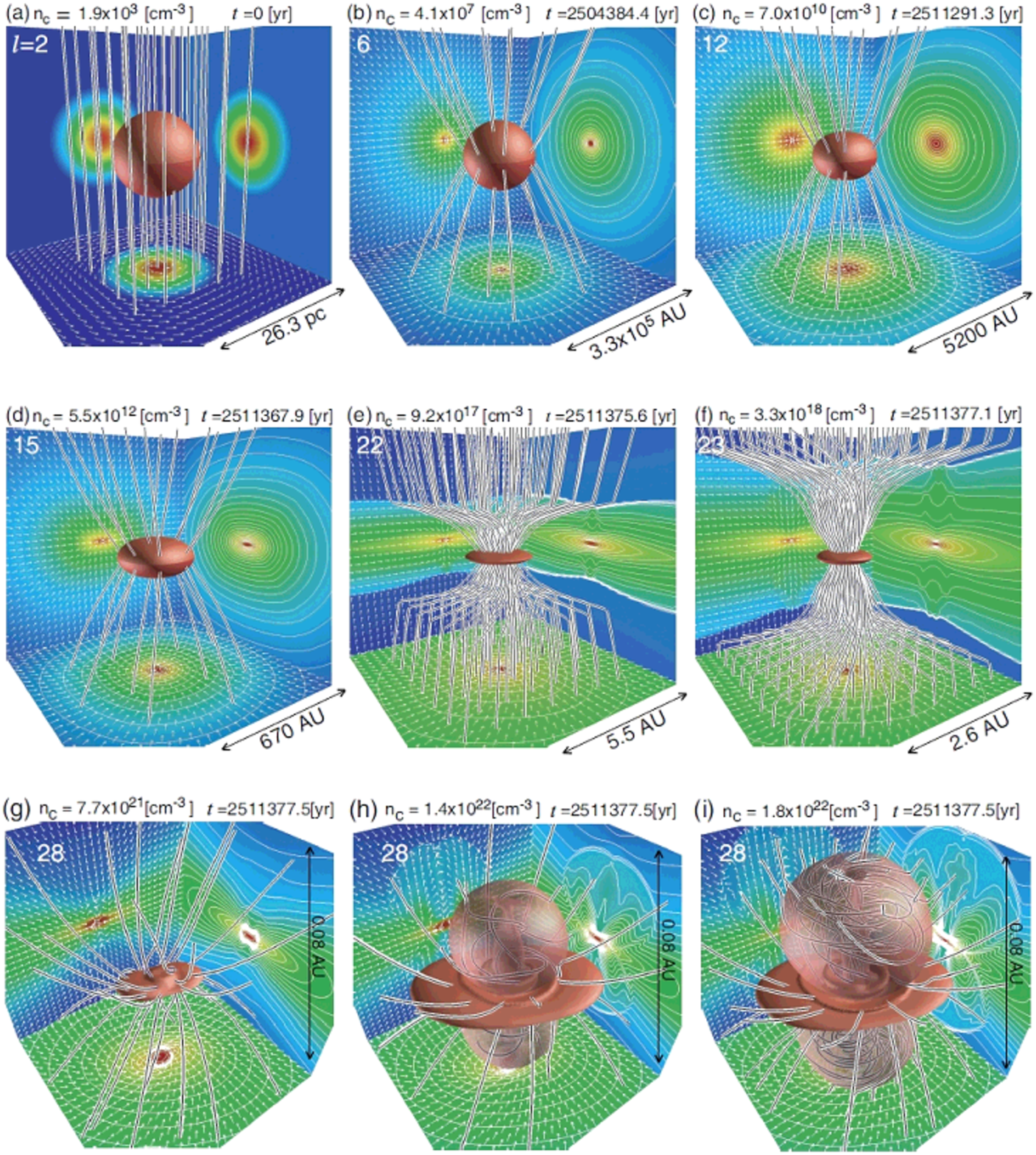}
\end{center}
\caption{
({\it a}--{\it i}) Time sequence of magnetically dominated model [($\beta_0$, $\gamma_0$) = ($10^{-4}$, $2\times10^{-3}$)] before protostar formation $\nc < 10^{21}\cm$.
In each panel, the structures of the high-density region ({\it isosurface}) and magnetic field lines ({\it black and white streamlines}) are plotted in three dimensions, while the density contours ({\it color} and {\it contour lines}) and velocity vectors ({\it thin arrows}) are projected on each wall surface.
The central number density $\nc$, elapsed time $t$, grid level $l$, and grid size are also shown in each panel.
The transparent red surface in panels ({\it g})-({\it i}) means the border between inflow and outflow inside which the mass ejected by the protostellar jet.
}
\label{fig:1}
\end{figure}

Figure~\ref{fig:1} shows the cloud evolution before protostar formation ($\nc<10^{21}\cm$) from the initial stage  for magnetically-dominated model that has a larger magnetic energy than the rotation energy ($\gamma_0 > \beta_0$).
This model has parameters of ($\gamma_0$, $\beta_0$) = ($2\times10^{-3}$, $10^{-4}$), and the initial cloud has a magnetic field strength of $B_0 = 10^{-6}$\,G, and an angular velocity of $\Omega_0=2.3\times10^{-16}$\,s$^{-1}$.
Figure~\ref{fig:1}{\it a} shows the initial spherical cloud threaded by a uniform magnetic field.
Figure~\ref{fig:1}{\it b}--{\it d} shows the cloud structure around the center of cloud when the central density reaches $\nc=$ ({\it b}) $4.1\times10^7$, ({\it c}) $7.0\times 10^{10}$, and ({\it d}) $5.5\times10^{12}\cm$, respectively.
The density contours projected on the sidewall in these figures indicate that the central region becomes oblate as the cloud collapses because of the magnetic field and rotation, both of which are amplified as the cloud collapses.
Figure~\ref{fig:1}{\it a}--{\it d} also shows that the magnetic field lines gradually converge toward the center as the central density increases.
Figure~\ref{fig:1}{\it g}--{\it i} shows the evolution after the protostar formation, in which the strong jet is driven by the circumstellar disk.
The speed of jet is $\sim100\km$ that corresponds to the Kepler speed near the protostar.

\begin{figure}
\begin{center}
\includegraphics[width=150mm]{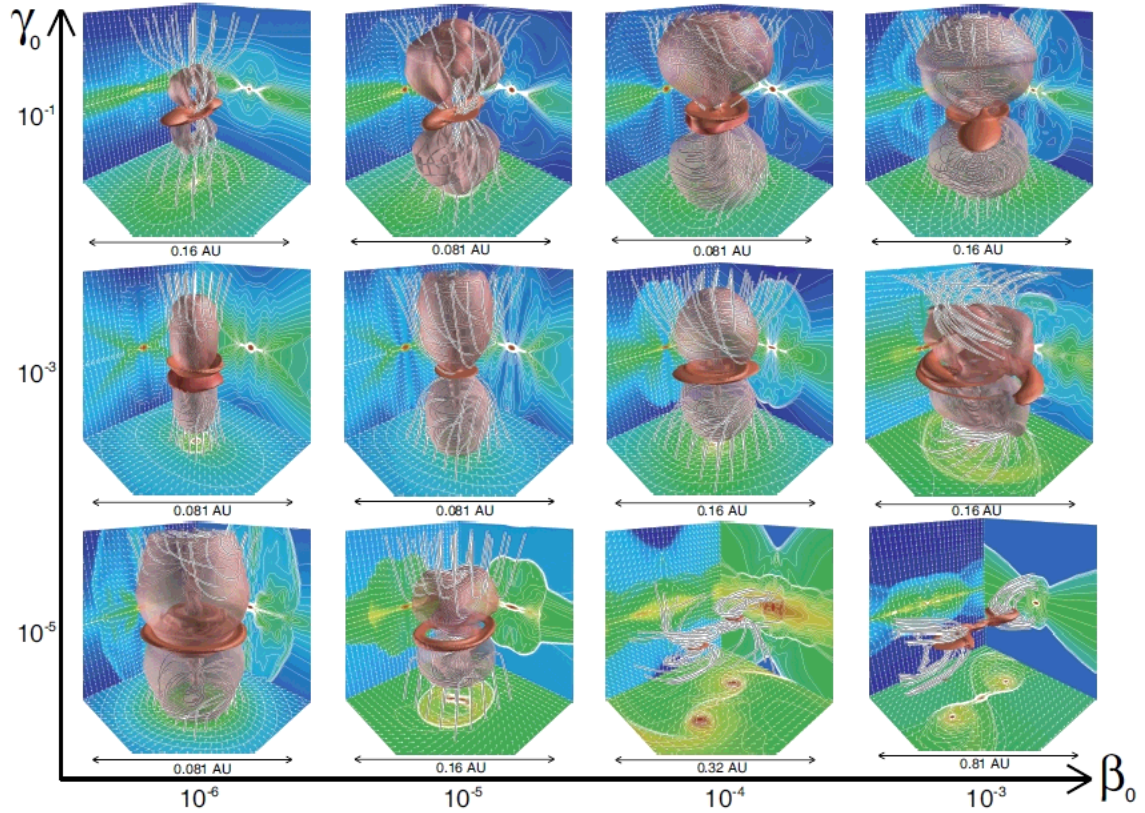}
\end{center}
\caption{
Final states in three dimensions against parameters $\beta_0$ and $\gamma_0$.
The magnetic field lines ({\it black and white streamlines}), high-density regions ({\it isosurface}), and jet ({\it transparent isosurface}) are plotted in each panel.
The density contours ({\it color} and {\it contour lines}) and velocity vectors ({\it thin arrows}) are also projected on each wall surface.
The grid level ($l$), and grid scale are shown in each panel.
}
\label{fig:2}
\end{figure}

Figure~\ref{fig:2} shows a configuration of magnetic field lines (black-white streamlines), a structure of a jet (transparent iso-velocity surface) and fragments for some models against $\beta_0$ - $\gamma_0$ plane.
The magnetically-dominated models ($\gamma_0 > \beta_0$) show a powerful jet after the proto-Population III star formation without fragmentation (or binary formation), while the rotation-dominated models ($\beta_0 > \gamma_0$) show fragmentation and binary formation without jet driving.
In the jet-driving models, the magnetic field lines are strongly twisted inside the jet region.
These jets have hourglass-like configurations of the magnetic field lines, where the poloidal component is more dominant than the toroidal component.
These configurations of the magnetic field lines can easily drive a strong jet by the disk wind mechanism.

In fragmentation models, the configurations of the magnetic field lines are disturbed, and the toroidal field tends to be more dominant than the poloidal filed.
Such configuration of the magnetic field lines means the weak magnetic field, which is insufficient to drive a jet.
The weak poloidal field is attributed to the oblique shocks on the surface of the disk envelope, and the disturbance of the field lines are attributed to the orbital motion of the fragments and the spin of the fragments.
The spin of the fragment winds the magnetic fields around its rotation axis, and the toroidal component of the magnetic field is amplified. 
Thus, in the further stages, the magnetic pressure may drive a jet.

\section{Summary}
In this study, we calculated cloud evolution from the stage of $n_c = 10^3\cm$ until the protostar is formed ($\simeq 10^{22}\cm$) for 36 models, parameterizing the initial magnetic field strength and rotation, to investigate effects of magnetic fields in collapsing primordial clouds.
Our calculations showed that fragmentation occurs but no jet appears when $\beta_0 > \gamma_0$, and jet appears after the protostar formation without fragmentation when $\beta_0 < \gamma_0$.
Thus, in the collapsing primordial cloud, the cloud evolution is mainly controlled by the centrifugal force than the Lorentz force when $\beta_0 > \gamma_0$, while the Lorenz force is more dominant than the centrifugal force when $\gamma_0  > \beta_0$.

A jet is driven when the initial cloud has magnetic field of 
\begin{equation}
B_0 \gtrsim 10^{-9} \left( \dfrac{\nc}{10^3\cm} \right)^{2/3} \,G  
\label{eq:bcrit}
\end{equation}
if the cloud rotates slowly as $\Omega \lesssim 4\times 10^{-17}(\nc/10^3\cm)^{2/3}$\,s$^{-1}$.
The power of a jet, e.g., a mass ejection rate, is considered to be controlled by the accretion rate as indicated in present-day star formation; the mass ejection rate of a jet is 1/10 of the mass accretion onto the protostar.
The accretion rate of primordial star formation is expected to be considerably larger than that at present day, and it produces a stronger jet.
The life time of the jet also seems to be controlled by accretion in the present-day; a jet stops when mass accretion stops.
For Population III stars, the gas accretion does not halt within their lifetimes \citep{omukai01,omukai03}.
Therefore, a jet also may continue during the all lifetime of the protostar, and the strong jet propagates to disturb a surrounding medium significantly.
The disturbance of the medium could trigger the subsequent star formation as frequently observed in present-day star formation.

Assuming the power law growth of $B_0 \propto n_c^{2/3}$, the critical strength of the magnetic field, $B_0 = 10^{-9}$\,G at $n_c = 10^{3}\cm$ corresponds to $B_0 = 5\times 10^{-13}$\,G at $n_c = 0.01\cm$ [see, eq.~(\ref{eq:bcrit})], which is much stronger than the background magnetic field  derived by \citet{ichiki06}.
However, when the magnetic field is amplified to $B \sim 10^{-9}(\nc/10^3\cm)^{2/3}$\,G by some mechanisms \citep[e.g.,][]{schleicher10}, the magnetic field can affect the collapse of the primordial cloud.
Even if a cloud has a magnetic field weaker than the critical strength $B_0 = 10^{-9}$\,G, the magnetic field may play an important role after the protostar formation.
In analytical study of the evolution of accretion disks around the first stars \citep{tan04} , it is suggested  that magnetic fields amplified in the circumstellar disk eventually give rise to protostellar jets during the protostellar accretion phase.

Rotation promotes fragmentation when the first collapsed objects has the angular velocity of $\Omega_0 \gtrsim 10^{-17}(\nc/10^3\cm)^{2/3}$\,s$^{-1}$.
The fragmentation is expected to produces binary or multiple stellar system.
When a multiple stellar system is formed, some stars can be ejected by close encounters.
At the protostar formation epoch, the protostar has a mass of $M\simeq 10^{-3}\msun$.
The ejected proto-Population III stars may evolve to metal-free brown dwarfs or low-mass stars.
When a binary component in a multiple stellar system is ejected from the parent cloud by protostellar interaction, a low-mass metal free binary may also appear in the early universe. 
It is considered that the extremely metal-poor ([Fe/H]$<$-5) stars \citep{christlieb01,frebel05} are formed as binary members from metal-free gas, and then have been polluted by the companion stars during the stellar evolution \citep{suda04}.
In addition, a binary frequency in Population III star may be comparable to or larger than that at present day \citep{komiya06,machida08c,machida08d,machida09a,machida09b}.
In order to confirm the ejection scenario, the further long-term calculations are required.


\end{document}